\documentstyle[11pt,aaspp4]{article}

\newcommand{\NH}{\mbox{$N_{\rm H}$}}        
\newcommand{\hi}{H{\small ~I}}
\newcommand{\hii}{H{\small ~II}}
\newcommand{\Msun}{$M_{\odot}$}

\begin{document}
\input{rotate}
 
\title{DETECTION OF COMPACT NUCLEAR X-RAY EMISSION IN NGC~4736}
\author{W.~Cui\altaffilmark{1}, D.~Feldkun\altaffilmark{1}, and 
R.~Braun\altaffilmark{2}}
\altaffiltext{1}{Massechusetts Institute of Technology, Center for Space Research, Cambridge, MA\ \ 02139}
\altaffiltext{2}{NFRA, Postbus 2, 7990 AA Dwingeloo, The Netherlands}
\authoremail{cui@space.mit.edu}
\begin{abstract}

  We report the results from a deep {\it ROSAT} PSPC observation of LINER 
galaxy NGC~4736. Two bright sources are detected, separated by only 
about 1\arcmin, with the brighter one coinciding with the center of the 
galaxy. Neither source shows
apparent X-ray variability on time-scales of minutes to hours in the 
{\it ROSAT} band. Simple power-law models, typical of AGN X-ray spectra, 
produce poor fits to the observed X-ray spectrum of the nuclear source. 
The addition of a Raymond-Smith component improves the fits significantly.
This is consistent with the presence of hot gas in the nuclear region with 
$kT\simeq 0.3$ keV, in addition to a compact nuclear source. 
However, a careful examination of the residuals reveal apparent features at 
low energies ($< 0.25$ keV). We find that the addition of a narrow emission
line at about 0.22 keV is a significant improvement to the parameterization
of the spectrum. We examine the results in the light of the accuracy of
the PSPC spectral calibration.
The derived photon index is about 2.3, which is similar to
those for Seyfert~1 galaxies measured in the {\it ROSAT} energy range. On the
other hand, the 0.1-2 keV luminosity for the compact source is 
only about $3.4\times 10^{39}\mbox{ }erg \mbox{ }s^{-1}$, much fainter 
than typical Seyfert galaxies. We discuss the implications of these
results on the connection between LINERs and AGNs.

  The off-center source is transient in nature. It has a hard X-ray spectrum, 
with a photon index of about 1.5, so is likely to be an X-ray binary. 
There is still some ambiguity regarding its association with the galaxy. If 
it is indeed located in the galaxy, the 0.1-2 keV luminosity would be greater
than $5.1\times 10^{38}\mbox{ }erg\mbox{ }s^{-1}$, making it a stellar-mass 
black hole candidate.

\end{abstract}


\section{Introduction}
  The active galactic nucleus (AGN) phenomenon expresses itself very
clearly at X-ray wavelengths. X-ray emission from AGN is thought to be
due to the accretion of surrounding matter onto a massive central black
hole. The X-ray luminosities of AGNs span many decades. Quasars are the
most luminous among the class, with X-ray luminosities above $10^{45}
erg\mbox{ }s^{-1}$. Seyfert galaxies, on the other hand, are much fainter, 
with typical luminosities of $10^{43-45} erg\mbox{ }s^{-1}$. However the 
least luminous ones can be as faint as $10^{40-41} erg\mbox{ }s^{-1}$ 
\markcite{cui96b}(Cui et al 1996b), which is
already close to the luminosity of bright ``normal'' galaxies 
\markcite{fabbiano95}(Fabbiano 1995). Only about 1\% of luminous galaxies are 
AGNs in a classical sense \markcite{weedman77}(Weedman 1977). It is therefore 
natural to ask if AGN activity stops
abruptly at Seyfert galaxies, or extends to lower luminosities, more
specifically, to the realm of ``normal'' galaxies.

  X-ray observations of nearby galaxies sometimes show relatively
bright nuclear emission \markcite{fabbiano89}(Fabbiano 1989), similar to 
that detected in
Seyfert galaxies. Optical spectroscopic surveys of these galaxies show
that as many as 10--20\% of the bright ones may be active, but
at a much lower level than classical AGNs. Many of them
have optical spectra characteristic of low ionization emission-line
regions (LINERs) (\markcite{heckman80}Heckman 1980; 
\markcite{staufer82}Staufer 1982; \markcite{keel83a}\markcite{keel83b}Keel 
1983a, b; \markcite{filippenko85}Filippenko \& Sargent 1985; 
\markcite{veron86}V$\acute{e}$ron-Cetty \& V$\acute{e}$ron 1986;
\markcite{phillips86}Phillips et al. 1986). These surveys also show 
that LINERs are very
common among ellipticals and early-type spirals, about 80\% of Sa
galaxies, over 40\% of Sb galaxies, and one third of all spiral
galaxies. Perhaps they are the missing link between Seyfert and ``normal''
galaxies.

  LINERs seem to make up a heterogeneous class: while some show
AGN-like activity, the observed nuclear activity in others may be
explained by other physical processes, such as shock heating, cooling
flow, or photoionization by hot stars at the center of the galaxy,
without invoking an AGN scenario (see a review by \markcite{filippenko94}
Filippenko 1994). 
Systematic investigations of AGN-like activity in LINERs have been 
carried out primarily in the optical (e.g., \markcite{filippenko92}
Filippenko \& Terlevich 1992; \markcite{ho93}Ho, Filippenko, \& Sargent 1993),
while observations at
higher energies, such as UV and X-rays, have only begun recently 
(\markcite{maoz95}Maoz et al. 1995; \markcite{koratkar95}Koratkar et al. 1995).

  To address this issue, we studied a nearby Sab galaxy NGC~4736 
\markcite{sandage74}(Sandage \& Tammann 1974). NGC~4736 has the optical 
spectrum of a LINER (\markcite{heckmanetal80}Heckman, Balick, \& Crane 1980; 
\markcite{heckman80}Heckman 1980), and is known optically
for its bright nucleus and high surface brightness in compact spiral
arms. Its optical similarities to the Seyfert galaxy NGC~1068 make it
``the best candidate for an ex-Seyfert'' \markcite{keel78}(Keel \& Weedman 
1978). Recently, \markcite{maoz95}Maoz et al. (1995) reported the detection 
of compact nuclear
UV emission in NGC~4736 using the Faint Object Camera on the {\it
Hubble Space Telescope} (HST), which suggests the presence of an
AGN-like continuum source or a compact stellar cluster 
at the center of this galaxy.

  In this paper, we present the results from a deep {\it ROSAT PSPC} 
observation of NGC~4736. The X-ray and radio data are described and
presented in \S~2. Detailed analyses are carried out in \S~3, and the
results discussed in \S~4. Finally, we conclude by summarizing the main 
results in \S~5.

\section{Data}
  NGC 4736, along with a few other high latitude, nearby, face-on
spiral galaxies, was chosen to be observed by the {\it ROSAT PSPC} as part of
a systematic investigation of the distribution of soft diffuse X-ray
emission, thus the distribution of hot gas, in the disks of ``normal''
spiral galaxies (\markcite{cui94}Cui 1994; \markcite{cui96a}Cui et al. 1996a).
Some basic data on NGC~4736 are summarized in Table~1.

  During the observation (RP600050), the pointing direction ($RA(J2000)=
12^h 50^m 50.4^s$, $DEC(J2000) = +41^d 06^m 36.0^s$) is slightly off
the center of the galaxy (see Table~1). The total on-source time is about 
96~ks. Sporadic short term enhancements of unknown origin \markcite{snowden94}
(Snowden et al. 1994) dominated the total count rate for a significant 
fraction of the observation. The data were also 
severely contaminated by scattered solar X-rays. While it is essential to
understand and remove these non-cosmic X-ray events for studying
diffuse X-ray emission from this galaxy \markcite{cui96a}(Cui et al. 1996a), 
it is not
as critical for the study of individual bright sources because the
conventional ``annulus background subtraction'' technique can be applied. 
However, in order to minimize possible systematic effects, such as 
non-uniform distribution of non-cosmic events over the detector, we chose 
to exclude the time intervals when the total count rate was 
20 $count\mbox{ }s^{-1}$ higher than the
``quiescent'' level. This eliminated short-term enhancements and the worst
instances of scattered solar X-rays; it leaves us with an exposure
time of about 87~ks. The X-ray image (with a pixel size of 7.5\arcsec) 
was re-projected it to epoch B1950, and was interpolated to a smaller 
pixel size. The final image of the central region of the galaxy is shown 
in Fig.~1, for the 0.1-2.0 keV energy band.

  In addition to the X-ray data, we have also obtained interferometric
\hi\ 21~cm data \markcite{braun95}(Braun 1995) and JCMT CO (2$\rightarrow$1) 
data for
this galaxy. A full discussion and analysis of these data will be
presented elsewhere. In this paper, we use these distributions
primarily to constrain the internal absorption of soft X-rays by gas
within the galaxy.

\section{Analysis}
  The X-ray image clearly shows two closely-spaced sources near the
center of the galaxy. They are only separated by about 1\arcmin, with
the brighter one coinciding with the center of the galaxy
(within the {\it ROSAT} pointing accuracy, usually $\sim$~10\arcsec).

\subsection{Spatial Analysis}
  To estimate the spatial extent of the detected sources, we made linear
cuts through them along both east-west and south-north directions, with
7.5\arcsec\ $\times$ 7.5\arcsec\  square bins. Fig.~2 
shows the linear profiles of both sources. Note that only one profile for 
the off-center source is shown because of heavy ``contamination'' by
the central source along the other direction. The solid-lines are the 
broad-band averaged theoretical point spread functions (PSFs) at the 
locations of
the sources. They fit the measured profiles well around the peaks,
indicating that the sources are consistent with being point-like. Broad
wings on the source profiles are apparent for both sources, and are
caused by the difficulty of separating them, as well as by the 
diffuse X-ray emission which is very strong at the center and hardly 
detectable in the background annulus \markcite{cui96a}(Cui et al. 1996a). 
For such a long exposure, the source position can be accurately determined
by fitting the source profile with PSF (to an accuracy much better than the 
spatial resolution of the {\it PSPC}). The best determined positions of the 
sources are listed in Table~2. In this case, errors are dominated by the 
{\it ROSAT} pointing uncertainty.

\subsection{Timing and Spectral Analysis}
  Because two sources are so close to each other, we chose to extract source 
counts within a pie-shaped region (see Table~2) centered on 
one source to avoid ``contamination'' from the other. The background counts 
were obtained from an annulus centered on the galactic center with radii of 
3.3\arcmin\  and 7.5\arcmin, which is nearly free of any other bright sources 
(only one was detected and removed).

  Neither source displays apparent X-ray variability on timescales of 
minutes to hours. 

  For each source the total counts were accumulated into a 256-channel, 
pulse-height-invariant spectrum. We ignored energy channels below 9 and 
above 212, and rebinned the remaining channels into 26 energy bins using the 
scheme adopted by the {\it ROSAT} Standard Analysis Software System. 

  For the central source, the high signal-to-noise ratio (S/N) of the 
resulting X-ray spectrum made it particularly challenging to adequately 
model it because systematic uncertainties in the instrument calibration 
become significant in this case. We tried to fit the spectrum with simple 
models, such as blackbody (BB), power-law (PL), and thermal bremsstrahlung 
(TB), using \markcite{morrison83}Morrison \& McCammon cross-sections (1983) 
for absorption, but none of them individually provided an adequate fit. 
Significant improvement was achieved with the addition of a Raymond-Smith 
component (RS; \markcite{raymond77}Raymond \& Smith 1977) which approximates
the X-ray emission from an optically-thin hot plasma. This is consistent 
with the reported 
detection of strong diffuse X-ray emission in the inner disk of NGC~4736 
\markcite{cui96a}(Cui et al. 1996a), because a likely origin of such emission 
is the 
thermal emission from million-degree gas present in the central region 
of the galaxy. As discussed above, the diffuse component should remain in 
the present data. Therefore, at least one RS component is required to 
adequately model it. A combination of PL (or TB or BB) and RS is thus the 
simplest plausible model to fit the observed X-ray spectrum, with PL (or TB
or BB) representing the spectrum of the compact source itself. Depending on 
the geometries and positions of the X-ray emitting regions in the galaxy, 
these two components may be absorbed by different amount of matter along the
line-of-sight. We proceeded to fit the spectra with such two-component models,
\begin{equation}
I_{observed} = I_{s} e^{-\sigma N_H^{s}} + I_{RS} e^{-\sigma N_H^{RS}},
\end{equation}
where s is either PL, TB, or BB, and assuming solar abundances for RS.
Both PL+RS and TB+RS fit the data equally well, and the best-fit parameters 
are shown in Table~3. Clearly, neither fit is statistically acceptable. 
The problem can be better demonstrated by Fig.~3a, which
shows the observed X-ray spectrum of the central source, and the best-fit
PL+RS model. Upon examination of the residual, the model deviates from
the data very significantly at the low energy portion of the spectra ($< 0.25$ 
keV), even though it fits the high energy portion reasonable well. 
We are not certain if this is a real effect, or calibration artifacts,
which would be more prominent in high $S/N$ data (we have used the 92mar11 
response matrix, as instructed in the {\it ROSAT} Users' Handbook). 
Spectral calibration uncertainties are known to be larger at lower energies 
(see the {\it ROSAT} Users' Handbook). As recommended, a 2\% systematic 
uncertainty was added to the data to represent the best estimated uncertainty
in detector response matrix calibration.

  In order to model the low-energy end of the spectrum, we added a soft BB 
component to the two-component models. Unfortunately, no significant 
improvement was obtained. Decent fits were made possible with the addition
of a Gaussian line feature (centered at $E_c\simeq 0.22$ keV with zero width; 
no absorption was applied to this component) to the two-component models. 
This feature might be associated with one or more of emission lines 
of highly ionized Fe, Si, Mg, and S in an optically thin plasma. We then 
experimented with adjusting abundances of these elements in the RS model
and in the absorption cross section calculation in an attempt to remove 
it, but failed to get a reasonable fit. This seems to suggest that the 
Gaussian component is a calibration artifact. The best-fit model (PL+RS+GAU) 
is plotted in Fig.~3b, and the parameters are shown in Table~3. 

  We experimented with other more complicated models that have been proposed 
for AGNs, including the partial-covering models that seem to fit the observed 
AGN X-ray spectra well \markcite{ceballos96}(Ceballos \& Barcons 1996). 
However, none provided 
significant improvement over the simple models. The large discrepancies 
between the models and the observed spectrum at energies below $\sim$0.25 keV 
are hardly reduced by adopting these complicated models.

   Similar spectral analysis was carried out for the off-center source.
Fig.~5 shows the observed X-ray spectrum for this source, along with the
best-fit model (PL+RS). In this case, no Gaussian line feature was required 
to model the observed X-ray spectrum (perhaps due to relatively low $S/N$ of 
the data). The results are summarized in Table~3.

\section{Discussion}
  As expected \markcite{cui96a}(Cui et al. 1996a), the diffuse component is 
dominated by the 1/4 keV emission, as evidenced by the temperature of the RS 
component (about 0.3 keV). It accounts for about 30-35\% of the total observed 
0.1-2.0 keV flux. The \NH\  values are not very well constrained in our case, 
although small values seem to be preferred. The radio data show, however, 
that although the 
integrated \hi\ emission shows a decline in {\it apparent} atomic column 
density inward of about 40\arcsec\  radius from about N$_{HI} = 2.\times 
10^{21}~cm^{-2}$ to less than about $10^{20}~cm^{-2}$ (see Fig.~1a and 
Fig.~4), there are indications that the \hi\ at radii less than about 
60\arcsec\  is already quite opaque so that {\it actual} \hi\ columns are 
likely to be several times larger. Comparable column densities of molecular 
hydrogen are inferred from CO emission at radii of about 50\arcsec. These 
appear to increase roughly exponentially to smaller radii (as shown in Fig.~4 
utilizing the data of
\markcite{gerin91}Gerin et al. 1991 and assuming a CO 1-0 to \NH\  
conversion factor of
$2.3\times 10^{20}\mbox{ }cm^{-2}/K\mbox{ }km\mbox{ }s^{-1}$). Although
some authors have postulated a ``hole'' in the molecular gas
distribution at small radii \markcite{smith91}(Smith et al. 1991), there is 
as yet no 
observational evidence for such a configuration of the neutral gas.
How can we reconcile these seemingly conflicting results? There are
two possibilities: (1) the \NH\  distribution in NGC~4736 is clumpy on 
angular 
scales smaller than the beam sizes of the \hi\  and CO surveys, and there is 
a relatively clear path toward the nuclear source, or (2) the detected 
source is located off the galactic nucleus, and sits high above molecular 
clouds in the region. Our results alone cannot distinguish between these 
scenarios. 

  We cannot rule out a purely thermal origin of the central source. 
Thus, the shock-heating scenario is still very likely. Hot gas is definitely 
present in the region, and produces strong diffuse X-ray emission 
\markcite{cui96a}(Cui et al.
1996a). On the other hand, a power-law X-ray source at the galactic 
center is also possible to account for the bulk of the emission. The derived 
power-law photon index for the source is similar to those for typical
Seyfert~1 galaxies measured in the {\it ROSAT} energy range (\markcite{turner93}Turner, George, \& Mushotzky 1993). With an assumed distance of 4 Mpc, the 
observed 0.1-2 keV luminosity of the source
is only about $3.4\times 10^{39} erg\mbox{ }s^{-1}$, much lower than that of
typical Seyfert galaxies. Under AGN settings, the low X-ray luminosity is 
perhaps due to an intrinsically lower mass accretion rate in the galaxy due 
to a lack of ``fuel'' in the region. Assuming an X-ray conversion efficiency 
of 10\%, a lower limit on the mass accretion rate can be estimated as
\begin{equation}
\mbox{\.{M}} \geq \frac{10L_{0.1-2.0keV}}{c^2},
\end{equation}
with $L_{0.1-2.0keV}\simeq 3.4\times 10^{39} erg\mbox{ }s^{-1}$ (see 
Table~3). 
The lower limit is thus $\sim 6\times 10^{-7}M_{\odot}\mbox{ }yr^{-1}$, 
which does not appear to be in conflict with any other observational lower 
limits on AGN activity. Another possibility is 
that the accretion process in this galaxy operates in a so-called 
``advection-dominated regime'' in which the bulk of released energy 
is advected with the gas present in the region into the central black hole;
little is radiated away in the form of X-rays (e.g., \markcite{narayan95a}
Narayan \& Yi 1995). 
This model could provide a natural explanation for the presence of hot gas in 
the region. The advection-dominated
accretion may be common among low-luminosity Seyfert galaxies and LINERs,
in fact, it appears to be what takes place at the center of our own
Galaxy, based on a multi-wavelength spectral analysis of Sagitarrius
$A^{*}$ \markcite{narayan95b}(Narayan, Yi, \& Mahadevan 1995).

  The nucleus of NGC~4736 shows signs of being active: it is bright in
H$\alpha$, as shown in Fig.~6 \markcite{pogge89}(Pogge 1989), and radio 
continuum emission (see Fig.~1b). \markcite{maoz95}Maoz et al. (1995) 
reported the detection of compact nuclear UV
emission in NGC~4736 in a {\it HST} observation. In fact, they detect
two unresolved sources of comparable brightness separated by about 
2.5\arcsec. The source surrounded with diffuse emission is presumed 
to coincide with the optical nucleus of NGC~4736, while the second 
source is suggested to be the nucleus of a recently merged sub-system.
With the unprecedented spatial resolution of the {\it HST},
this detection provides evidence for the existence of a non-stellar (or
extremely compact stellar) continuum source. We have now detected 
luminous compact X-ray emission from the region, and the measured 
position of the central X-ray source is consistent with it being located 
at the galactic nucleus (see Tables 1 and 2). However, because the spatial 
resolution of the {\it ROSAT PSPC} is poor, we cannot rule out the 
possibility that the observed X-ray emission may be the integrated 
contribution from luminous X-ray binaries near the galactic center. 

   Compact nuclear X-ray emission has also been detected in other LINERs. 
\markcite{koratkar95}Koratkar et al. (1995) studied five low-luminosity 
AGNs (LLAGNs) including two LINERs to address specifically the issue of 
whether the 
dominant X-ray production mechanism is the same at all luminosities in AGNs. 
Soft X-ray emission was detected in all galaxies. The high-resolution 
{\it HRI} images show that the emission is mostly or entirely confined to
the nucleus. The observed {\it ROSAT PSPC} spectra of the LINERs and one 
Seyfert~1 galaxy in their sample are similar, and can be fit by simple 
power-law models with relatively low internal absorption. Like NGC~4736, 
the derived photon indices and absorbing column densities are similar 
to those for luminous Seyfert~1 galaxies measured in the {\it ROSAT} energy 
range (\markcite{turner93}Turner, George, \& Mushotzky 1993).
Evidence is strong to support that the physical processes in producing 
X-rays in LLAGNs (including a significant fraction of LINERs) are the same 
as in the luminous objects, although studies of a larger sample of LLAGNs 
are clearly required to draw any definitive conclusions.

   What is the nature of the off-center source? The observed 0.1-2.0 keV
luminosity (internally absorbed) is $\sim 5\times 10^{38} erg\mbox{ }s^{-1}$, 
making it a very luminous
X-ray source, if it is indeed a single point source. The hard power-law
spectrum seems to suggest that it is likely to be a X-ray binary. The
mass of the primary in the binary system can then be estimated as
\begin{equation} 
M=M_{\odot} \frac{L}{L_E},
\end{equation}
where $L_E$
is Eddington luminosity for a 1\Msun\  accreting object. Assuming an
Eddington luminosity appropriate for pure electron scattering, $\sim
1.3\times 10^{38} erg\mbox{ }s^{-1}$, the {\it absorbed} X-ray luminosity 
(see Table~3) would require
the mass to be larger than $\sim$ 4\Msun, making it a stellar-mass black hole
candidate. Similar luminous X-ray sources have also been detected in several 
other nearby spiral galaxies (see \markcite{fabbiano95}Fabbiano 1995 for a 
review).

   This source is located on the inner \hi\  ring, to the north-west of
the nuclear source. Many compact radio sources were detected along the
ring, with both thermal and non-thermal radio continuum spectra 
\markcite{duric88}(Duric
\& Dittmar 1988). They are thought to be thermal bremsstrahlung
emission from compact \hii\  regions associated with hot, young stars,
and synchrotron radiation from young supernova remnants respectively,
both of which indicate active star formation in the ring. High star
formation rates are also supported by the observed H$\alpha$ emission
in the region \markcite{pogge89}(Pogge 1989). The detected X-ray source 
is only a few
arcseconds away from one of the thermal radio sources, as shown in 
Fig.~6, which is associated with a compact \hii\  region. The poor 
spatial resolution of the {\it ROSAT PSPC} prevents a definitive
identification with the compact radio source.

   The proximity of this source to the nuclear source might also
suggest some causal relation: is it some sort of ejecta or jet from the
central source. The observed X-ray morphology, as shown in Fig.~1,
seems to argue against it, but high spatial resolution is clearly
needed to address this question.

   It is also conceivable that what we are seeing are the cores of two merged 
galaxies, both of which emit strongly in X-rays. This model may provide
a natural explanation for the formation of the prominent ring structures in 
this galaxy \markcite{mulder93}(Mulder \& Van Driel 1993), since the 
disturbance caused by the galaxy merging process is likely to produce
strong shock waves. These same shocks might also trigger star-formation 
in the rings, as supported by radio and H$\alpha$ observations 
(\markcite{duric89}Duric \& Dittmar 1989; \markcite{pogge89}Pogge 1989). 
In this case, the observed off-center UV source (\markcite{maoz95}Maoz et al. 
1995) may simply be a circumnuclear star cluster.

   However, the derived \NH\  values are uncomfortably low (lower than
that due to the foreground interstellar medium in our Galaxy; see Table~1). 
When we fixed the \NH\  to the Galactic value, the fits became worse.
This seems to indicate that the source is closer to us than the galaxy, 
even though the 
probability of a foreground X-ray source fortuitously lining up with
the nucleus of the galaxy is very small. There are, however, no known
Galactic X-ray sources in this direction. Of course, it may be a transient 
source that went into a relatively long and steady outburst state when 
observed by the PSPC, similar to 4U1630-47 (a black hole candidate), as 
currently seen by the All-Sky Monitor on the {\it Rossi X-ray Timing 
Explorer}. When we were about to submit this paper, the {\it ROSAT} HRI image
of NGC~4736 became available (RH600678, which was carried out roughly 4 
years after the PSPC observation). As expected, the HRI image shows a bright 
X-ray source at the galactic center, but the off-center source had disappeared
completely, which supports its transient nature. Instead, another source
is detected about the same distance away from the nucleus, but to the west
and slightly south. (We have carefully checked the orientation of the
observation by using the bright X-ray sources in the field). Is this 
consistent with the proper motion of a nearby star? If so, the star would
have moved an angular distance of $\sim$25\arcsec\  in $\sim$4 years, which
would imply a distance (of the source from us) less than $\sim$ 13 pc,
assuming a maximum tangential velocity of $400\mbox{ }km\mbox{ }s^{-1}$. As 
a result, the observed X-ray luminosity would be less than 
$\sim 5\times 10^{27}\mbox{ }erg\mbox{ }s^{-1}$, which is still consistent 
with stars with X-ray emitting coronae (e.g., \markcite{schmitt90}Schmitt et 
al. 1990). It cannot be a flare star, however, because no strong X-ray 
variability was measured. We have also searched, and have not found any 
reported recent supernovae in NGC~4736. It is more likely, therefore, that 
both X-ray sources are associated with the galaxy itself, as supported by 
their excellent alignment with the nuclear source. This would provide strong 
evidence that the inner \hi\ ring (at which both sources are located) is 
indeed quite active. 

\section{Conclusions}
   We have presented and discussed the results of our analysis of a
deep {\it ROSAT} observation of the nearby galaxy NGC~4736 to
investigate possible AGN-like activity in its nuclear region, as
suggested by studies of this galaxy at other wavelengths. Our main
results can be summarized as follows: 

\begin{description} 
\item[$\bullet$] Two bright X-ray sources are detected in the central
region of NGC~4736. Their X-ray radial profiles are consistent with the
PSFs of the {\it ROSAT PSPC}, and the separation is estimated to be
about 1 kpc. The brighter source is consistent with being at the
nucleus, suggesting that it may perhaps represent the non-stellar
continuum source that powers the observed nuclear emission at various
wavelengths. 
\item[$\bullet$] The detailed spectral analysis show that
the observed X-ray emission from the nuclear region can be attributed
both to a central source and hot gas present in the region, with the
central source contributing the bulk of the observed 0.1-2.0 keV 
luminosity. The X-ray spectrum of the off-center source is harder. 
It is likely to be a luminous, transient X-ray binary. Then, the measured 
X-ray luminosity would make it a stellar-mass black hole candidate.
\end{description}

   Combining our results with those from observations of NGC~4736
at other wavelengths, particularly the {\it HST} observation, evidence is
strong for the presence of a non-stellar continuum source at the nucleus
of the galaxy, similar to those thought to exist in Seyfert galaxies. 

\acknowledgments
We wish to thank an anonymous referee for useful comments that resulted in an
improved manuscript. This research has made use of the archival database 
maintained by the High 
Energy Astrophysics Science Archive Research Center (HEASARC) at the NASA 
Goddard Space Flight Center, and the NASA/IPAC Extragalactic Database
(NED) which is operated by the Jet Propulsion Laboratory, Caltech, under
contract with the National Aeronautics and Space Administration. This work
is supported in part by NASA Contract NAS5-30612.

\clearpage

\begin{deluxetable}{lll}
\tablecolumns{3}
\tablewidth{0pc}
\tablecaption{Basic Data on NGC 4736}
\tablehead{\colhead{Name} & \colhead{Value} & \colhead{References}}
\startdata
Type			& Sab		& \markcite{sandage77}Sandage \& Tammann 1987 \nl
RA (J2000)		& $12^h$ $50^m$ $53^s$	& \markcite{sandage77}Sandage \& Tammann 1987 \nl
DEC (J2000)		& $41^d$ $07^m$ $17^s$	& \markcite{sandage77}Sandage \& Tammann 1987 \nl
Galactic \NH\  ($10^{20} cm^{-2}$) & 1.4 & \markcite{dickey90}Dickey \& Lockman 1990 \nl
Distance		& 4 Mpc	& e.g. \markcite{duric88}Duric \& Dittmar 1989 \nl
Major Diameter		& 14\arcmin	& \markcite{nilson73}Nilson 1973 \nl
Minor Diameter		& 12\arcmin	& \markcite{nilson73}Nilson 1973 \nl
\hi/\hii\  ring		& 40\arcsec-60\arcsec	& \markcite{mulder93}Mulder \& Van Driel 1993\nl
Outer \hi\  ring	& 4\arcmin-6\arcmin 	& \markcite{mulder93}Mulder \& Van Driel 1993\nl
Distance scale 	& $\sim 19pc/1\arcsec$ & assuming a distance of 4 Mpc\\
\enddata
\end{deluxetable}

\clearpage

\begin{deluxetable}{lll}
\tablecolumns{3}
\tablewidth{0pc}
\tablecaption{The Detected Sources}
\tablehead{\colhead{Quantity} & \colhead{Nuclear Source} & \colhead{Off-center Source}}
\startdata
RA (J2000)\tablenotemark{1} & $12^h$ $50^m$ $52^s$ & $12^h$ $50^m$ $48^s$ \nl 
DEC (J2000)\tablenotemark{1} & $+41^d$ $07^m$ $10^s$ & $+41^d$ $07^m$ $35^s$ \nl
Remaining Exposure Time (ks) & 87 & 87 \nl
Extracted Source Counts & 17432 & 1432 \nl
Extracted Fraction\tablenotemark{2}& 80\% & 50\% \nl
Hardness Ratio\tablenotemark{3}& $0.026\pm 0.008$  & $0.135\pm 0.026$ \nl
\tablenotetext{1}{Errors (not shown) are dominated by the {\it ROSAT} pointing 
uncertainty ($\sim$10\arcsec; see text).}
\tablenotetext{2}{Fractional area of the pie-shaped region in a circle.}
\tablenotetext{3}{Hardness ratio is defined as $(H-S)/(H+S)$, where H is the 
total counts in energy channels 41-200, and S in channels 8-40.}
\enddata
\end{deluxetable}

\clearpage

\vskip 2.0cm

\begin{table}[p]
\newbox\rotbox
\setbox\rotbox = \hbox{
\renewcommand{\thempfootnote}{\alph{mpfootnote}}
\renewcommand{\thefootnote}{\alph{footnote}}
\renewcommand{\footnoterule}{}
\begin{minipage}{8.5in}
\begin{center}
\small
  \caption{Results of Spectral Analysis$^1$}
  \vspace{0.3cm}
  {\renewcommand{\arraystretch}{1.5}
  \begin{tabular}{llccccccccccc}  \hline \hline
Source & Model & $N_H^{s}$ & $\alpha$ & $kT^{TB}$ & $N_H^{RS}$ & $kT^{RS}$ & Line Energy & Line Width & $\chi^2/dof$ &$F^{2}$ & $L^{2}$ & $f_{s}^{2}$ \\
 & &$10^{20} cm^{-2}$ & & $keV$&$10^{20} cm^{-2}$ & $keV$ & $keV$ & $keV$ & & & & \% \\ \hline
Nuclear & PL+RS+GAU & $6^{+10}_{-3}$ & $2.3^{+0.5}_{-0.4}$ &$-$&$3^{+20}_{-2}$& $0.31^{+0.04}_{-0.16}$ & $0.224^{+0.003}_{-0.004}$ & $0$ & $54.6/21$ & $17.8$ & $3.4$ & 58\\
  & PL+RS & $2.0$ & $2.5$ & $-$ & $1.4$ & $0.42$ & $-$ & $-$ & $124.3/24$ &$-$ & $-$ & $-$ \\
  & TB+RS+GAU &$4^{+6}_{-2}$ & $-$ & $1.5^{+2.2}_{-0.4}$ & $5^{+58}_{-4}$ & $0.29^{+0.03}_{-0.15}$ & $0.225^{+0.002}_{-0.003}$ & $0$ & $56.9/21$ & $17.2$ & $3.2$ & 54\\
  & TB+RS &$1.3$ & $-$ & $0.7$ & $1.2$ & $0.44$ & $-$ & $-$ & $182.0/24$ & $-$ & $-$ & $-$\\
Off-center &PL+RS& $0.4^{+0.4}_{-0.4}$ & $1.5^{+0.4}_{-0.2}$ &$-$&$48^{+133}_{-48}$ & $0.4^{+0.4}_{-0.2}$ & $-$ & $-$ & $32.9/24$ & $2.62$ & $0.51$ &76\\
 & &$1.4$ (fixed) & $1.8^{+0.1}_{-0.1}$ &$-$& $1.4$ (fixed) & $0.7^{+0.3}_{-0.2}$ & $-$ & $-$ & $46.4/26$ & $-$ & $-$ &$-$\\
 &TB+RS&$0.2^{+0.4}_{-0.2}$&$-$&$4^{+16}_{-2}$ &$46^{+66}_{-46}$&$0.4^{+0.4}_{-0.3}$ & $-$ & $-$ &$33.4/24$&$2.64$&$0.51$ &79\\
 & &$1.4$ (fixed)&$-$&$1.3^{+0.2}_{-0.3}$&$1.4$ (fixed)&$0.7^{+0.7}_{-0.4}$& $-$ & $-$ &$55.3/26$&$-$&$-$ &$-$\\ \hline
\end{tabular} }
\end{center}
\vspace{0.1cm}
\noindent $^{1}$ Uncertainties represent the 90\% confidence interval.

\noindent $^{2}$ F: the total observed 0.1-2.0 keV flux in 
units of $10^{-13} erg\mbox{ }s^{-1} cm^{-2}$; L: the corresponding 
X-ray luminosity in units of $10^{39}erg\mbox{ }s^{-1}$, calculated for an 
assumed distance of 4 Mpc; and $f_{s}$: the fractional source (PL or TB)
contribution. Flux and Luminosity have been corrected for insufficient 
source-count extraction (see discussion in \S~2). 
\end{minipage}
} \rotl\rotbox
\end{table}

\newpage
\figcaption[fig.1]{X-ray image of the central region of NGC~4736 in 
the 0.1-2.0 keV energy band (in units of source counts), overlaid with (a)
contours of apparent H~I column density from 7.5 to 20.0 with a step 
size of \mbox{$2.5\times 10^{20} cm^{-2}$}, and (b) contours of 21 cm radio
continuum emission with brightness 2.0, 2.8, 4.0, 5.6, 8.0, 
11.2, and 16.0 mJy/beam (the beam size is 15\arcsec). Note 
that the X-ray peak is arbitrarily aligned with the radio continuum 
peak with a 10.9\arcsec\  shift in RA and 2.6\arcsec\  shift in DEC, which 
is within the {\it ROSAT} pointing uncertainty. }

\figcaption[fig.2]{Linear profiles of the nuclear source along
(a) the east-west and (b) north-south directions, and (c) the off-center 
source along the north-south direction, with a square pixel size of 
7.5\arcsec. 
The solid lines are the broad-band averaged theorectical PSFs 
at these locations. The broad wings are perhaps due to diffuse X-ray 
emission in the region, as well as due to the difficulty of separating
two sources (see text).}

\figcaption[fig.3]{X-ray spectrum of the nuclear source. The 
solid-line histograms represent the best-fit models to the data: (a) PL+RS 
and (b) PL+RS+GAU. }

\figcaption[fig.4]{Radial profiles of the {\it apparent} absorbing column 
density due to atomic and molecular neutral hydrogen in NGC~4736 (see text).}

\figcaption[fig.5]{X-ray spectrum of the off-center source. The 
solid-line histogram represents the best-fit PL+RS model to the data.}

\figcaption[fig.6]{Overlay of the detected X-ray sources in color on the 
image of compact radio continuum sources detected in the region, along with 
H$\alpha$ contours, taken from Duric \& Dittmar (1988). The X-ray image 
was smoothed with a 52.5\arcsec\  $\times$ 52.5\arcsec\  square box filter. 
Note that the X-ray image is shifted by the same amount as in Fig.~1. }


\clearpage

\begin{references}
\reference{braun95}Braun,~R. 1995, \aaps, 114, 409
\reference{ceballos96}Ceballos,~M.~T., \& Barcons,~X. 1996, \mnras, in press
\reference{cui94}Cui,~W. 1994, Ph.D. Thesis, University of Wisconsin-Madison
\reference{cui96a}Cui,~W., Sanders,~W.~T., McCammon,~D., Snowden,~S.~L., \& Womble, ~D.~S. 1996a, \apj, 468, 102
\reference{cui96b}Cui,~W., et al. 1996b, in preparation
\reference{dickey90}Dickey,~J.~M., \& Lockman,~F.~J. 1990, \araa, 28, 215
\reference{duric88}Duric,~N., \& Dittmar,~M.~R. 1988, \apjl, 332, L67
\reference{fabbiano95}Fabbiano,~G 1995, in X-ray Binaries, ed. W.~H.~G.~Lewin, J.~van Paradijs, \& E.~P.~J.~van den Heuvel (Cambridge: Cambridge University Press), 390.
\reference{fabbiano89}Fabbiano,~G 1989, \araa, 27, 87
\reference{filippenko85}Filippenko,~A.~V., \& Sargent,~W.~L.~W. 1985, \apjs, 57, 503
\reference{filippenko92}Filippenko,~A.~V., \& Terlevich,~R. 1992, \apjl, 397, L79
\reference{filippenko94}Filippenko,~A.~V. 1994, in The Nearest Active Galaxies, ed. J.~E.~Beckman, L.~Colina, \& H.~Netzer (Madrid: CSIC), 259
\reference{gerin91}Gerin,~M., Casoli, F., \& Combes, F. 1991 \aap, 251, 32
\reference{heckman80}Heckman,~T.~M. 1980, \aap, 87, 152
\reference{heckmanetal80}Heckman,~T.~M., Balick,~B., \& Crane,~P.~C. 1980, \aaps, 40, 295
\reference{ho93}Ho,~L.~C., Filippenko,~A.~V., \& Sargent,~W.~L.~W. 1993, \apj, 417, 63
\reference{keel83a}Keel,~W.~C. 1983a, \apj, 269, 466
\reference{keel83b}Keel,~W.~C. 1983b, \apjs, 52, 229
\reference{keel78}Keel,~W.~C., \& Weedman,~D.~W. 1978, \aj, 83, 1
\reference{koratkar95}Koratkar,~K., Deustua,~S.~E., Heckman,~T., Filippenko,~A.~V., Ho,~L.~C., \& Rao,~M. 1995, \apj, 440, 132
\reference{maoz95}Maoz,~D., Filippenko,~A.~V., Ho,~L.~C., Rix,~H., \& Bahcall,~J.~N. 1995, \apj, 440, 91
\reference{morrison83}Morrison,~R., \& McCammon,~D. 1983, \apj, 270, 119
\reference{mulder93}Mulder, P.S., \& Van Driel, W. 1993, \aap, 272, 63
\reference{narayan95a}Narayan,~R., \& Yi,~I. 1995, \apj, 453, 480
\reference{narayan95b}Narayan,~R., Yi,~I., \& Mahadevan,~R. 1995, Nature, 374, 623
\reference{nilson73}Nilson,~P. 1973, Uppsala General Catalog of Galaxies
\reference{phillips86}Phillips,~M.~M., Jenkins,~C.~R., Dopita,~M.~A., Sadler,~E.~M., \& Binette,~L. 1986, \aj, 91, 1062
\reference{pogge}Pogge,~R.~W. 1989, \apjs, 71, 433
\reference{raymond77}Raymond,~J.~C., \& Smith,~B.~W. 1977, ApJS, 35, 419; updated by Raymond,~J.~C. 1991 (adopted in XSPEC package)
\reference{sandage87}Sandage,~A., \& Tammann,~G.~H. 1987, A revised Shapley-Ames Catalog of Bright Galaxies (Washington D. C: Carnegie Institute of Washington)
\reference{schmitt90}Schmitt,J.~H.~M.~M.~, Collura,~A., Schortino,~S., Vaiana,~G.~S., Harnden,~Jr,~F.~R., \& Rosner,~R. 1990, \apj, 365, 704
\reference{smith91}Smith, B.J., Lester, D.F., Harvey, P.M., \& Pogge, R.W. 1991, \apj, 373, 66
\reference{snowden94}Snowden,~S.~L., McCammon,~D., Burrow,~D.~N., \& Mendenhall,~J.~A. 1994, \apj, 424, 714
\reference{stauffer82}Stauffer,~J.~R. 1982, \apj, 262, 66
\reference{turner93}Turner,~T.~J., George,~I.~M., \& Mushotzky,~R.~F. 1993, \apj, 412, 72
\reference{veron86}V$\acute{e}$ron-Cetty,~M.~P., \& V$\acute{e}$ron,~P. 1986, \aaps, 66, 335
\reference{weedman77}Weedman,~D.~W. 1977, \araa, 15, 69
\end{references}
\end{document}